# Impact of (SARS-CoV-2) COVID 19 on the indigenous language-speaking population in Mexico

**Authors**
Carlos Medel-Ramírez[1], Hilario Medel-López[2]

**Affiliations**
1. Universidad Veracruzana / Instituto de Investigaciones y Estudios Superiores Económicos y Sociales
2. Universidad Veracruzana / Instituto de Antropología

**Corresponding author**
Carlos Medel-Ramírez ( cmedel@uv.mx )

**Abstract**
The importance of the working document is that it allows the analysis of the information and the status of cases associated with (SARS-CoV-2) COVID-19 as open data at the municipal, state and national level, with a daily record of patients, according to a age, sex, comorbidities, for the condition of (SARS-CoV-2) COVID-19 according to the following characteristics: a) Positive, b) Negative, c) Suspicious. Likewise, it presents information related to the identification of an outpatient and / or hospitalized patient, attending to their medical development, identifying: a) Recovered, b) Deaths and c) Active, in Phase 3 and Phase 4, in the five main population areas speaker of indigenous language in the State of Veracruz - Mexico. The data analysis is carried out through the application of a data mining algorithm, which provides the information, fast and timely, required for the estimation of Medical Care Scenarios of (SARS-CoV-2) COVID-19, as well as for know the impact on the indigenous language-speaking population in Mexico.

**Specifications Table**

| Subject | Infectious Diseases |
|---|---|
| Specific subject area | Information from the Viral Respiratory Diseases Epidemiological Surveillance System for (SARS-CoV-2) COVID-19 in Mexico corresponding to the indigenous language-speaking population in Mexico. |
| Type of data | Table<br>Figure |

| How data were acquired | Government of Mexico. Health Secretary. Databases Covid-19 México https://datos.gob.mx/busca/dataset/informacion-referente-a-casos-covid-19-en-mexico/resource/e8c7079c-dc2a-4b6e-8035-08042ed37165 <br> Instruments: <br> Software Orange Data Mining version 3.26.0 https://orange.biolab.si <br> Make and model and of the instruments used: <br> Algorithm for the identification of patients according to following characteristics: a) Positive, b) Negatives, c) Suspects. Likewise, it presents information regarding the identification of an outpatient and / or hospitalized patient, attending to their medical development, identifying: a) Recovered, b) Deaths and c) Assets |
|---|---|
| Data format | The information is presented in raw in CVS format, the Ministry of Health of Mexico since April 14, 2020 published the cases associated with (SARS-CoV-2) COVID-19 as open data. The data processing corresponds to the records on the epidemic (SARS-CoV-2) COVID-19 at 1 August 2020. The treatment of the information is carried out through the application software for data mining Orange version 3.26.0, in which the algorithm for the analysis of information is filtered to present the current scenario of the indigenous language-speaking population in Mexico of the SARS-CoV-2 (COVID 19). |
| Parameters for data collection | The information is presented at the municipal, state and national levels, with a daily registry of patients, according to age, sex, comorbidities, for the condition of (SARS-CoV-2) COVID-19 according to the following characteristics: a) Positive, b) Negatives, c) Suspects. Likewise, it presents information regarding the identification of an outpatient and / or hospitalized patient, attending to their medical development, identifying: a) Recovered, b) Deaths and c) Assets. |
| Study area | The states that comprise the indigenous language speaking population in Mexico are: Aguascalientes … Zacatecas |
| Description of data collection | This information is filtered to present the current scenario in the indigenous language speaking population in Mexico of the SARS-CoV-2 (COVID 19) in a fast and timely manner, to support public decision-making in health matters. |
| Data source location | Institution: Universidad Veracruzana / Instituto de Investigaciones y Estudios Superiores Económicos y Sociales <br> Country: México |
| Data accessibility | Raw data can be retrieved from the Github repository https://github.com/CMedelR/dataCovid19/edit/master/README.md |

## Value of the Data

- The Algorithm for the identification of patients (SARS-CoV-2) COVID 19 in Mexico allows to analyze at the municipal, state and national level, the registry of patients, according to age, sex, comorbities, for condition of (SARS-CoV-2) COVID-19 according to the following characteristics: a) Positive, b) Negative, c) Suspicious, as well as presenting information on the identification of an outpatient and / or hospitalized patient, attending to their medical development, identifying: a) Recovered, b ) Deaths and c) Assets, in Phase 3 and Phase 4, in a fast and timely manner, to support public decision-making in health matters.
- Taking into account their strategic roles in public health and researchers can use the data from this study to identify the action scenario for decision-making in the combat of (SARS-CoV-2) COVID 19 in Phase 3 and Phase 4 corresponding to the indigenous language speaking population in Mexico
- The importance of data analysis is that it allows identifying cases (SARS-CoV-2). COVID-19 in Mexico is concentrated daily and knowing the impact on the population and allows preparing action scenarios to make public health policy decisions to combat SARS-CoV-2) COVID-19 in the indigenous language speaking population in Mexico

**Data Description**

The source of information on the number of registered cases of (SARS-CoV-2) COVID-19 at 1 August 2020 for Mexico comes from the website https://datos.gob.mx/busca/dataset/informacion-referente-a-casos-covid-19-en-mexico by the Ministry of Health, with the participation of the National Council for Science and Technology (CONACYT), the Center for Research in Geospatial Information Sciences (CENTROGEO), the National Laboratory for Geo-Intelligence (GEOINT), the Data Laboratory of the National Laboratory for Geointelligence (DataLab), where the registry of COVID-19 cases (SARS-CoV-2) COVID-19 is concentrated, and is the official means of communication and information on the epidemic in the Popoluca from the Soteapan Area in Mexico.

The information of the cases (SARS-CoV-2) COVID-19 in Mexico is concentrated on a daily basis since April 19, 2020, communication and official information on the epidemic in Mexico, the data are presented at the municipal, state and national levels, with a daily registry of patients, according to age, sex, comorbities, for the condition of (SARS-CoV-2) COVID-19 according to the following characteristics: a) Positive, b) Negatives, c) Suspects. Likewise, it presents information regarding the identification of an outpatient and / or hospitalized patient, attending to their medical development, identifying: a) Recovered, b) Deaths and c) Assets. The data processing corresponds to the records on the epidemic (SARS-CoV-2) COVID-19 at 1 August 2020. The treatment of the information is carried out through the application software for data mining and visual programming Orange Data Mining version 3.26.0. Orange Data Mining is a machine learning and data mining suite for data analysis through Python scripting and visual programming. [1]

According to (WHO, 2020) the (SARS-CoV-2) COVID-19 disease pattern presents 4 scenarios identified from the confirmation of Laboratory Diagnosis: a) Not Infected or b) Infected, in this finally, the following categories are observed, taking into account age and specific comorbities in each case: a) Mild Infection, b) Moderate Infection, c) Severe Infection and d) Critical Infection.

Depending on the category observed in Patients who have a Confirmation of Infected, as in the case of a) or b) it can assume the character of Outpatient, so the strategy is isolation or "quarantine" at home, where the result It is hoped that he will recover. Regarding the Patients who have a Confirmation of Infected, in categories c) and d) they assume the character of Hospitalized Patient, with a probability of requiring care in Intensive Care Units and requiring Intubation, and where it is hoped to save as many patients as possible.

The importance of the research is that it allows identifying the action scenario for making public health policy decisions to combat CO(SARS-CoV-2) COVID-19, since they consider the following states of process in medical treatment, in order to carry out the Estimate of Scenarios for Medical Care of the (SARS-CoV-2) COVID-19 under the following premises of hospital care:

1. A patient with a positive (SARS-CoV-2) COVID-19 laboratory diagnosis can be considered: a) Outpatient, or b) Hospitalized.
2. If the (SARS-CoV-2) COVID-19 Positive patient is Hospitalized, the following should be considered: a) Enter the Intensive Care Unit or b) Do not enter the Intensive Care Unit.
3. If the (SARS-CoV-2) COVID-19 Positive patient is Hospitalized and Entered into the Intensive Care Unit, the following should be considered: a) The patient requires intubation or b) The patient does NOT require intubation.

**Methods**

The information is presented in raw in CVS format, the Ministry of Health of Mexico. The data processing corresponds to the records on the epidemic (SARS-CoV-2) COVID-19 at 1 August 2020. The treatment of the information is carried out through the application software for data mining Orange version 3.26.0, in which the algorithm for the analysis of information are developed and it is filtered to present the current scenario in Mexico of the SARS-CoV-2 (COVID 19). In this way, the algorithm that is presented allows us to project the requirements for the use of installed infrastructure in the face of the growing requirement for patient care Positive (SARS-CoV-2) COVID-19, allowing the identification of scenarios at the national, state and municipal levels. The construction of the algorithm is based on the following definitions.

**Definition 1:** Total Patients to consider in Model (SARS-CoV-2) COVID-19.- It is the number of total patients according to the confirmatory laboratory result or not of (SARS-CoV-2) COVID-19).
Be:
TP SARS-CoV-2 i j = Total patients according to (SARS-CoV-2) COVID-19 confirmatory laboratory result
Which consists of:
TP SARS-CoV-2 i j = (P + SARS-CoV-2 i j) + (P- SARS-CoV-2 i j) + (Px SARS-CoV-2 i j), where: i = State, j = Municipality
Of which:
P+ SARS-CoV-2 i j = Patient with a positive (SARS-CoV-2) COVID-19 result in the State, Municipality
P- SARS-CoV-2 i j = Patient with negative (SARS-CoV-2) COVID-19 result in the State, Municipality
Px SARS-CoV-2 i j = Patient with pending confirmation (SARS-CoV-2) COVID-19 in the State, Municipality

**Definition 2:** Identification of a suspected (SARS-CoV-2) COVID-19 case.- This is the patient who undergoes an initial qualification according to the initial diagnostic characteristics indicated in the case definitions for surveillance by the World Health Organization for primary care of (SARS-CoV-2) COVID-19 cases.
Be:
CsCOVID 19 (SARS-CoV-2) = Patient with initial classification as a suspected case of (SARS-CoV-2) COVID-19
Where:
Cs (SARS-CoV-2) COVID-19 = Cs (SARS-CoV-2) COVID-19 Type 1 + Cs (SARS-CoV-2) COVID-19 Type 2 + Cs (SARS-CoV-2) COVID-19 Type 3
Of which:
According to the World Health Organization, there are 3 categories (identified as Type 1, Type 2 and Type 3) to identify suspected cases of (SARS-CoV-2) COVID-19, defined below:
1. Cs (SARS-CoV-2) COVID-19 Type 1.- Is a patient with acute respiratory disease (fever and at least one sign / symptom of respiratory disease, with no other aetiology that fully explains the clinical presentation and a history of travel or residence in a country / area or territory that reports local transmission of COVID-19 disease during the 14 days prior to the onset of symptoms.

2. Cs (SARS-CoV-2) COVID-19 Type 2.- He is a patient with an acute respiratory disease, who has been in contact with a confirmed or probable COVID-19 case in the last 14 days before the onset of symptoms.
3. Cs (SARS-CoV-2) COVID-19 Type 3.- Is a patient with severe acute respiratory infection (fever and at least one sign / symptom of respiratory illness (eg cough, shortness of breath) and requiring hospitalization and without another etiology that fully explains the clinical presentation.

**Definition 3:** Total Patients to consider in the (SARS-CoV-2) COVID-19 Model .- It is the number of total patients according to the confirmatory laboratory result or not of (SARS-CoV-2) COVID-19).
Be:
TP SARS-CoV-2 i j = Total patients according to confirmatory laboratory result or not of (SARS-CoV-2) COVID-19
Which consists of:
TP SARS-CoV-2 i j = (P + SARS-CoV-2 i j) + (P- ARS-CoV-2 i j) + (Px ARS-CoV-2 i j) , where: i = State, j = Municipality
Of which:
P + SARS-CoV-2 i j = Patient with a positive (SARS-CoV-2) COVID-19 result in the State, Municipality
P- ARS-CoV-2 i j = Patient with negative (SARS-CoV-2) COVID-19 result in the State, Municipality
Px ARS-CoV-2 i j = Patient with pending confirmation (SARS-CoV-2) COVID-19 in the State, Municipality

**Definition 4:** Positive Patients for (SARS-CoV-2) COVID-19 i j.- It is the number of patients with laboratory results with positive confirmation for (SARS-CoV-2) COVID-19 i j .
It has:
P + SARS-CoV-2 i j = Patient with a positive (SARS-CoV-2) COVID-19 result in the State, Municipality

**Definition 5.-** Medical Treatment Strategy for a patient with positive laboratory confirmation for (SARS-CoV-2) COVID-19 i j .- It is the Action Plan in Medical Treatment for a patient with positive laboratory confirmation for SARS-CoV-2 in attention to your degree of infection and comorbidities present that is channeled to determine the Physician.

According to the Strategy of Medical Care required for Patients with a Positive SARS-CoV-2 Result, according to their degree of identified infection, they have the following.
Be:
ET P + SARS-CoV-2 i j = Medical Treatment Strategy P + SARS-CoV-2 i j

The medical treatment for a patient with a positive laboratory result for (SARS-CoV-2) COVID-19, based on the Medical Treatment Strategy (ETM P + SARS-CoV-2 ij), based on his degree of infection and present comorbidities, poses two action scenarios : i) Outpatient (SARS-CoV-2) COVID-19 patient or ii) Hospitalized (SARS-CoV-2) COVID-19 patient.
Be:
i) Outpatient COVID19 patient.
   P + SARS-CoV-2 i j Outpatient = Positive (SARS-CoV-2) COVID-19 with Outpatient mode in the State, Municipality
ii) COVID19 Patient Hospitalized.
   P + SARS-CoV-2 i j Hospitalized = Positive (SARS-CoV-2) COVID-19 with modality Hospitalized in the State, Municipality
where:
Depending on the degree of infection (I1, I2 or I3), the Hospitalized (SARS-CoV-2) COVID-19 Patient may require: i) Access to the Intensive Care Area without Intubation or ii) Access to the Intensive Care Area with Intubation.

**Definition 6.-** Patients with a Positive (SARS-CoV-2) COVID-19 Result Hospitalized with Access to the Intensive Care area.- It is the number of Patients with a Positive SARS-CoV-2 Result Hospitalized with Access to the Intensive Care area, according to its degree of infection.
Be:
P + SARS-CoV-2 i j Hospital Intensive Care = Positive (SARS-CoV-2) COVID-19 with modality Hospitalized in the State, Municipality

**Definition 7.-** Patients with a positive (SARS-CoV-2) COVID-19 result Hospitalized with access to the Intensive Care Area with Intubation.- It is the number of Patients with a Positive (SARS-CoV-2) COVID-19 Result Hospitalized with Access to the Intensive Care area with Intubation.
Be:
P + SARS-CoV-2 i j Hospital Intensive Care with Intubation = Positive (SARS-CoV-2) COVID-19 with Hospitalized modality and intubation in the State, Municipality.

**Definition 8.-** P + SARS-CoV-2 i j Deaths.- Deaths of Patients with a positive result for SARS-CoV-2. Deaths are all those positive to (SARS-CoV-2) COVID-19 where one is indicated in the data record (DATE_DEF other than the value "99-99-9999").

**Definition 9.-** (SARS-CoV-2) COVID-19 case fatality rate.- It is the proportion of people who die from (SARS-CoV-2) COVID-19 among the Patients with a positive (SARS-CoV-2) COVID-19 result in a given period and area.
Be:
TL SARS-CoV-2 i j = (SARS-CoV-2) COVID-19 case fatality rate
Where:
(SARS-CoV-2) COVID-19 case fatality rate = [(Deaths of Patients with a Positive (SARS-CoV-2) COVID-19 Result in the State or Municipality) / (Total of Patients with a Positive (SARS-CoV-2) COVID-19 result in the State or Municipality)] x 100
Of which:
DP+ SARS-CoV-2 i j = Deaths of Patients with a positive (SARS-CoV-2) COVID-19 result in the State / Municipality
And:
P+ SARS-CoV-2 i j = Total Patients with a positive (SARS-CoV-2) COVID-19 result in the State / Municipality
So:
TL SARS-CoV-2 i j = [D P + SARS-CoV-2 i j / P + SARS-CoV-2 i j] x 100

The data processing corresponds to the records on the epidemic (SARS-CoV-2) COVID-19 at 1 August 2020. The treatment of the information is carried out through the application software for data mining Orange version 3.26.0, in which the algorithm for the information analysis are developed. (See Figure 1, below). According to information from the Ministry of Health, the following records are available at the national level:

Figure 1. Algorithm for the identification of patients (SARS-CoV-2) COVID 19 in the indigenous language speaking population in Mexico Orange Data Mining version 3.26.0

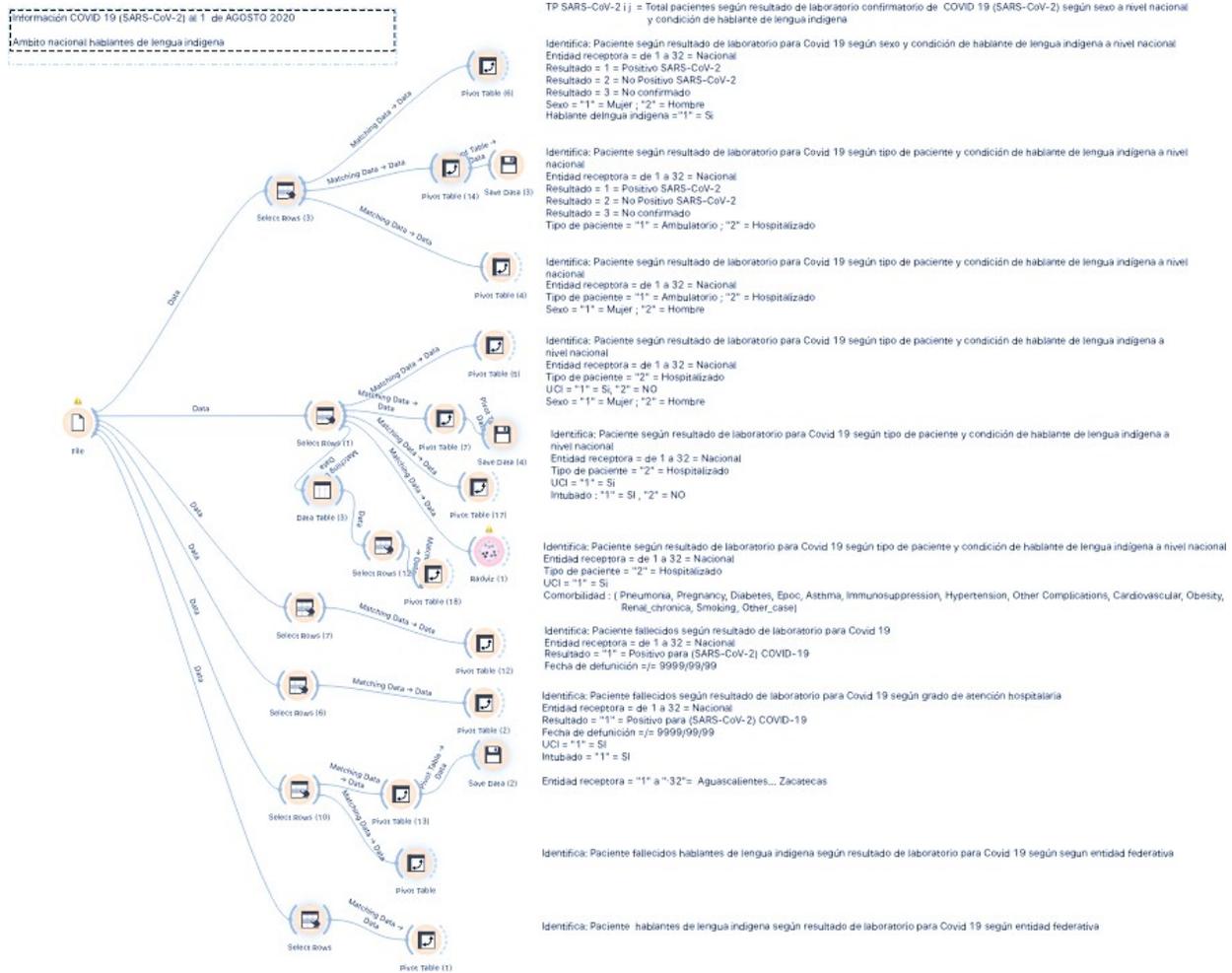

Source: Self made. With information provided by the Ministry of Health as of 1 June 2020 and Orange Data Mining version 3.25.1

**Working paper: Data mining for the study of the COVID-19 Epidemic (SARS-CoV-2) and the impact on the indigenous population in Veracruz: Algorithm for the identification of COVID patients (SARS-CoV-2), August to 1, 2020**

According to information from the Ministry of Health in Mexico, the following records are available at the national level:

1. The total number of cases in the indigenous language-speaking population in Mexico as of august 1, 2020, is 8,938 cases of which, based on the confirmatory or non-laboratory result for (SARS-CoV-2) COVID-19, the following classification is obtained: a ) 4,555 patients with a positive result for (SARS-CoV-2) COVID-19, b) 3,780 patients with a non-positive result for (SARS-CoV-2) COVID-19 and c) 603 patients with a pending result to determine (SARS-CoV -2) COVID-19. (See table 1)
2. The number of patients indigenous language-speaking population with a positive (SARS-CoV-2) COVID-19 result is 4,555 of which: a) 5,909 are care outpatients and b) 3,029 are hospitalized patients. (See table 2)
3. The total number of patients (SARS-CoV-2) COVID-19 hospitalized (include positive and non positive results) is 3,209 of whom 1,253 are women and 1,776 are men. (See Table 3)
4. The total number of hospitalized patients with a positive result for (SARS-CoV-2) COVID-19 is 1,830, of whom: a) 164 patients enter the intensive care unit; while b) 1,666 patients do not enter the intensive care unit. (See Table 4)
5. Only 84 Hospitalized with a positive (SARS-CoV-2) COVID-19 patients admitted to the intensive care unit required intubation; while 80 patients did not require intubation. (See Table 5)
6. Likewise, to date 3,486 deceased patients indigenous language-speaking population in Mexico with positive (SARS-CoV-2) COVID-19 have been registered, of which 67 deaths corresponded to Positive (SARS-CoV-2) COVID-19 patients who were in intensive care and intubation and 3,419 corresponded to (SARS-CoV-2) COVID-19 positive patients who were not in intensive care. (See Table 6).
7. The fatality rate of patients with a positive result for SARS-CoV-2) COVID-19 at nationwide as of august 1, 2020, is estimated at 10.93%. The five states that registered the highest fatality rates as of August 1, 2020 are the following: a) Morelos with a rate of 2039%, b) Baja California with 19.67%, c) Sinaloa with a rate of 17.46% , d) Colima with a rate of 17: 00% and e) State of Mexico with a rate of 15.37%. For its part, the State of Veracruz registered a rate of 13.20% while Mexico City registered a rate of 9.75% (See Table 7).
8. The fatality rate of patients indigenous language-speaking population in Mexico with a positive result for SARS-CoV-2) COVID-19 at nationwide as of august 1, 2020, is estimated at (See Table 8).
9. Figure 1 shows the main comorbidities identified in hospitalized patients indigenous language-speaking in Mexico with  a positive result for (SARS-CoV-2) COVID-19, nationwide as of August 1, 2020.
10. The Map 1 (SARS-CoV-2) COVID-19 fatality rate by State in Mexico as of August 1, 2020, while Map 2 shows the fatality rate for COVID-19 (SARS-CoV-2) in the indigenous-speaking population that required service in Intensive Care Units in Mexico by state as of August 1, 2020.

Tables and graphs

Table 1. Total number of cases in the indigenous language-speaking population in Mexico as of August 1, 2020, according to sex and result for (SARS-CoV-2) COVID-19

|  | Sex | | |
| --- | --- | --- | --- |
| Result | Women | Men | Total |
| Positive (SARS-CoV-2) COVID-19 | 1,845 | 2,710 | 4,555 |
| No positive (SARS-CoV-2) COVID-19 | 1,937 | 1,843 | 3,780 |
| Pending result | 286 | 317 | 603 |
| Total | 4,068 | 4,870 | 8,938 |

Source: Government of Mexico. Health Secretary. Information from the Epidemiological Surveillance System for Viral Respiratory Diseases 1 August 2020

Table 2. Total number of cases in the indigenous-speaking population in Mexico as of August 1, 2020, by type of hospital care

|  | Patient type | | |
| --- | --- | --- | --- |
| Result | Ambulatory | Hospitalized | Total |
| Positive (SARS-CoV-2) COVID-19 | 2,722 | 1,833 | 4,555 |
| No positive (SARS-CoV-2) COVID-19 | 2,757 | 1,023 | 3,780 |
| Pending result | 430 | 173 | 603 |
| Total | 5,909 | 3,029 | 8.938 |

Source: Government of Mexico. Health Secretary. Information from the Epidemiological Surveillance System for Viral Respiratory Diseases 1 August 2020

Table 2. Total number of cases in the indigenous-speaking population in Mexico as of August 1, 2020, according to sex and of hospital care

|  | Patient type | | |
| --- | --- | --- | --- |
| Sex | Ambulatory | Hospitalized | Total |
| Women | 2,815 | 1,253 | 4,068 |
| Men | 3,094 | 1,776 | 4,870 |
| Total | 5,909 | 3,029 | 8,938 |

Source: Government of Mexico. Health Secretary. Information from the Epidemiological Surveillance System for Viral Respiratory Diseases 1 August 2020

Table 4. Total number of cases in the indigenous-speaking population in Mexico as of August 1, 2020, according type of patient and Intensive care unit

| Patient type | Intensive care unit | No Intensive care unit | Total |
|---|---|---|---|
| Hospitalized | 164 | 1,666 | 1,830 |

Source: Government of Mexico. Health Secretary. Information from the Epidemiological Surveillance System for Viral Respiratory Diseases 1 August 2020

Table 5. Total number of cases in the indigenous-speaking population in Mexico as of August 1, 2020 according to hospitalized patient in intensive care unit and intubation condition

| Patient type | Intubated patient | No Intubated patient | Total |
|---|---|---|---|
| Hospitalized intensive care unit | 84 | 80 | 164 |

Source: Government of Mexico. Health Secretary. Information from the Epidemiological Surveillance System for Viral Respiratory Diseases 1 August 2020

Table 6. Total number of cases in the indigenous-speaking population in Mexico as of August 1, 2020 with positive COVID-19 (SARS-CoV-2), according to hospital care condition as of August 1, 2020

| Patient type | Women | Men | Total |
|---|---|---|---|
| At the national level | 1,085 | 2,401 | 3,486 |
| Intensive care unit patients with intubation | 18 | 49 | 67 |
| % Deceased patients requiring care in intensive and intubation, by sex | 1.66% | 2.04% | 1.92% |

Source: Government of Mexico. Health Secretary. Information from the Epidemiological Surveillance System for Viral Respiratory Diseases 1 August 2020

Table 7. (SARS-CoV-2) COVID-19 fatality rate by State in Mexico as of August 1, 2020

|  | Estado | Total deceased | Covid positives 19 | Case fatality rate |
|---|---|---|---|---|
| 1 | Aguascalientes | 257 | 4,120 | 6.24% |
| 2 | Baja California | 2,674 | 13,594 | 19.67% |
| 3 | Baja California Sur | 182 | 4,498 | 4.05% |
| 4 | Campeche | 500 | 4,588 | 10.90% |
| 5 | Chiapas | 645 | 13,116 | 4.92% |
| 6 | Chihuahua | 194 | 1,875 | 10.35% |
| 7 | Ciudad de México | 7,244 | 74,314 | 9.75% |
| 8 | Coahuila de Zaragoza | 967 | 5,775 | 16.74% |
| 9 | Colima | 904 | 5,317 | 17.00% |
| 10 | Durango | 288 | 4,167 | 6.91% |
| 11 | Guanajuato | 1,009 | 21,378 | 4.72% |
| 12 | Guerrero | 1,415 | 11,032 | 12.83% |
| 13 | Hidalgo | 1,033 | 6,901 | 14.97% |
| 14 | Jalisco | 1,549 | 13,313 | 11.64% |
| 15 | México | 8,225 | 53,513 | 15.37% |
| 16 | Michoacán de Ocampo | 780 | 9,910 | 7.87% |
| 17 | Morelos | 836 | 4,101 | 20.39% |
| 18 | Nayarit | 391 | 3,518 | 11.11% |
| 19 | Nuevo León | 1,118 | 18,032 | 6.20% |
| 20 | Oaxaca | 968 | 10,673 | 9.07% |
| 21 | Puebla | 2,446 | 20,355 | 12.02% |
| 22 | Querétaro | 449 | 3,726 | 12.05% |
| 23 | Quintana Roo | 1,015 | 7,840 | 12.95% |
| 24 | San Luis Potosí | 522 | 9,921 | 5.26% |
| 25 | Sinaloa | 2,230 | 12,774 | 17.46% |
| 26 | Sonora | 1,949 | 17,890 | 10.89% |
| 27 | Tabasco | 1,988 | 21,747 | 9.14% |
| 28 | Tamaulipas | 1,008 | 17,130 | 5.88% |
| 29 | Tlaxcala | 713 | 4,627 | 15.41% |
| 30 | Veracruz de Ignacio de la Llave | 2,849 | 21,582 | 13.20% |
| 31 | Yucatán | 875 | 10,098 | 8.67% |
| 32 | Zacatecas | 249 | 2,768 | 9.00% |
|  | Total | 47,472 | 434,193 | 10.93% |

Source: Own elaboration with Government of Information from the Mexico. Health Secretary. Epidemiological Surveillance System for Viral Respiratory Diseases as of August 1, 2020

Map 1 (SARS-CoV-2) COVID-19 fatality rate by State in Mexico as of August 1, 2020

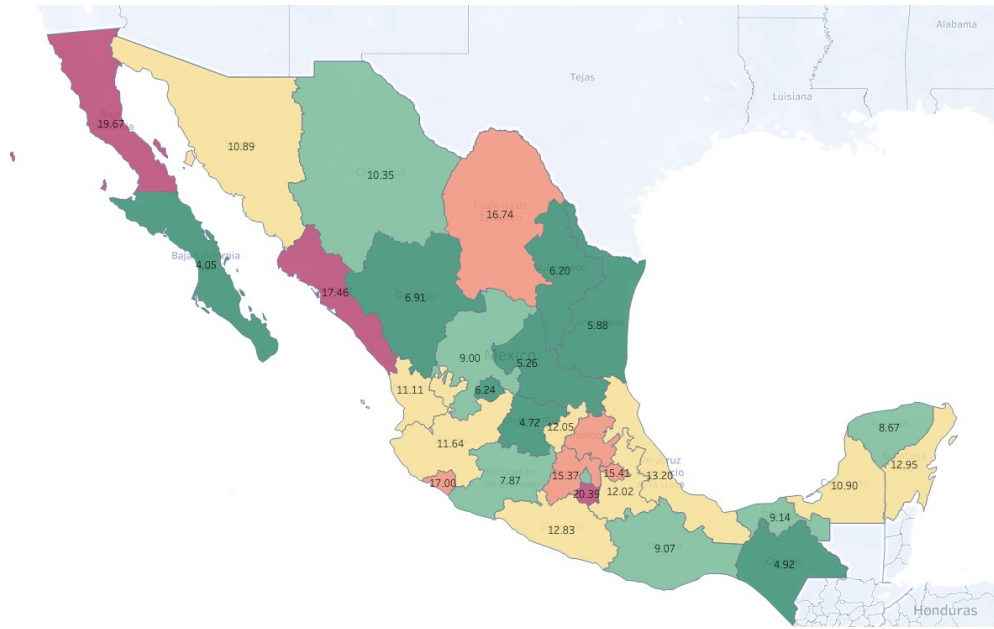

Source: Own elaboration with Government of Information from the Mexico. Health Secretary. Epidemiological Surveillance System for Viral Respiratory Diseases as of August 1, 2020

Table 8. (SARS-CoV-2) COVID-19 fatality rate on the indigenous language-speaking population in Mexico by State in Mexico as of August 1, 2020

|  | Estado | Total deceased | Covid positives 19 | Case fatality rate |
|---|---|---|---|---|
| 1 | Aguascalientes |  | 8 | 0.00 |
| 2 | Baja California |  | 81 | 0.00 |
| 3 | Baja California Sur |  | 6 | 0.00 |
| 4 | Campeche | 1 | 124 | 0.81 |
| 5 | Chiapas |  | 23 | 0.00 |
| 6 | Chihuahua |  | 7 | 0.00 |
| 7 | Ciudad de México | 5 | 149 | 3.36 |
| 8 | Coahuila de Zaragoza | 3 | 45 | 6.67 |
| 9 | Colima | 2 | 318 | 0.63 |
| 10 | Durango | 4 | 28 | 14.29 |
| 11 | Guanajuato |  | 45 | 0.00 |
| 12 | Guerrero | 5 | 228 | 2.19 |
| 13 | Hidalgo | 4 | 257 | 1.56 |
| 14 | Jalisco | 3 | 41 | 7.32 |
| 15 | México | 9 | 329 | 2.74 |
| 16 | Michoacán de Ocampo |  | 138 | 0.00 |
| 17 | Morelos | 1 | 38 | 2.63 |
| 18 | Nayarit |  | 54 | 0.00 |
| 19 | Nuevo León |  | 41 | 0.00 |
| 20 | Oaxaca | 18 | 410 | 4.39 |
| 21 | Puebla | 1 | 200 | 0.50 |
| 22 | Querétaro |  | 9 | 0.00 |
| 23 | Quintana Roo | 2 | 253 | 0.79 |
| 24 | San Luis Potosí |  | 276 | 0.00 |
| 25 | Sinaloa |  | 31 | 0.00 |
| 26 | Sonora | 1 | 116 | 0.86 |
| 27 | Tabasco |  | 127 | 0.00 |
| 28 | Tamaulipas |  | 19 | 0.00 |
| 29 | Tlaxcala |  | 30 | 0.00 |
| 30 | Veracruz de Ignacio de la Llave | 1 | 131 | 0.76 |
| 31 | Yucatán | 7 | 979 | 0.72 |
| 32 | Zacatecas |  | 14 | 0.00 |
|  | Total | 67 | 4,555 | 1.47 |

Source: Own elaboration with Government of Information from the Mexico. Health Secretary. Epidemiological Surveillance System for Viral Respiratory Diseases as of August 1, 2020

Map 2 (SARS-CoV-2) COVID-19 fatality rate on n the indigenous-speaking population that required service in Intensive Care Units in Mexico by state as of August 1, 2020

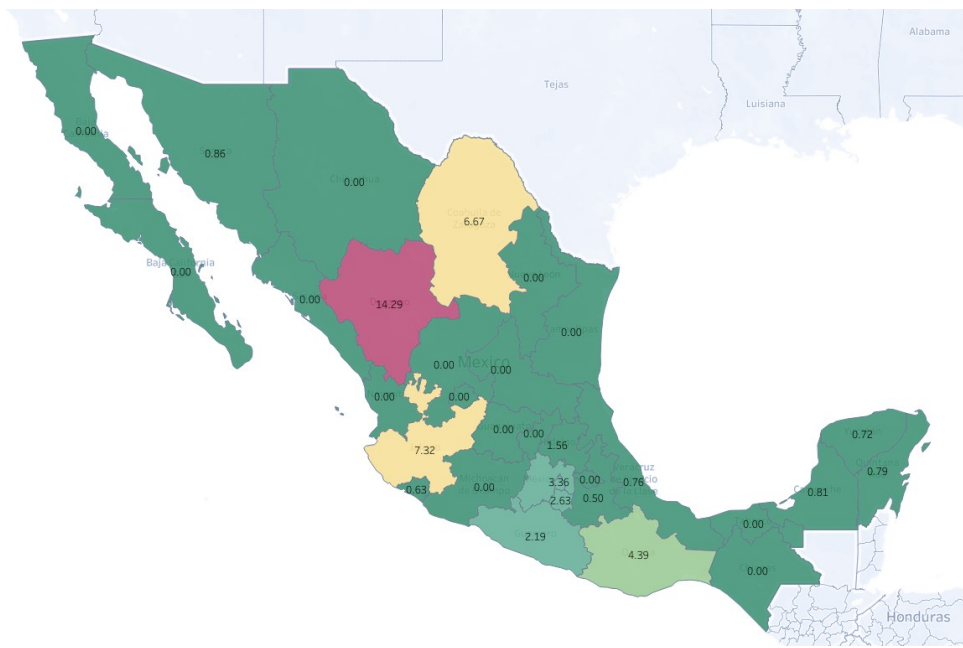

Source: Own elaboration with Government of Information from the Mexico. Health Secretary. Epidemiological Surveillance System for Viral Respiratory Diseases as of August 1, 2020

Figure 1 Comorbidities identified in hospitalized Indigenous-language-speaking patients with a positive result for (SARS-CoV-2) COVID-19, nationally as of August 1, 2020

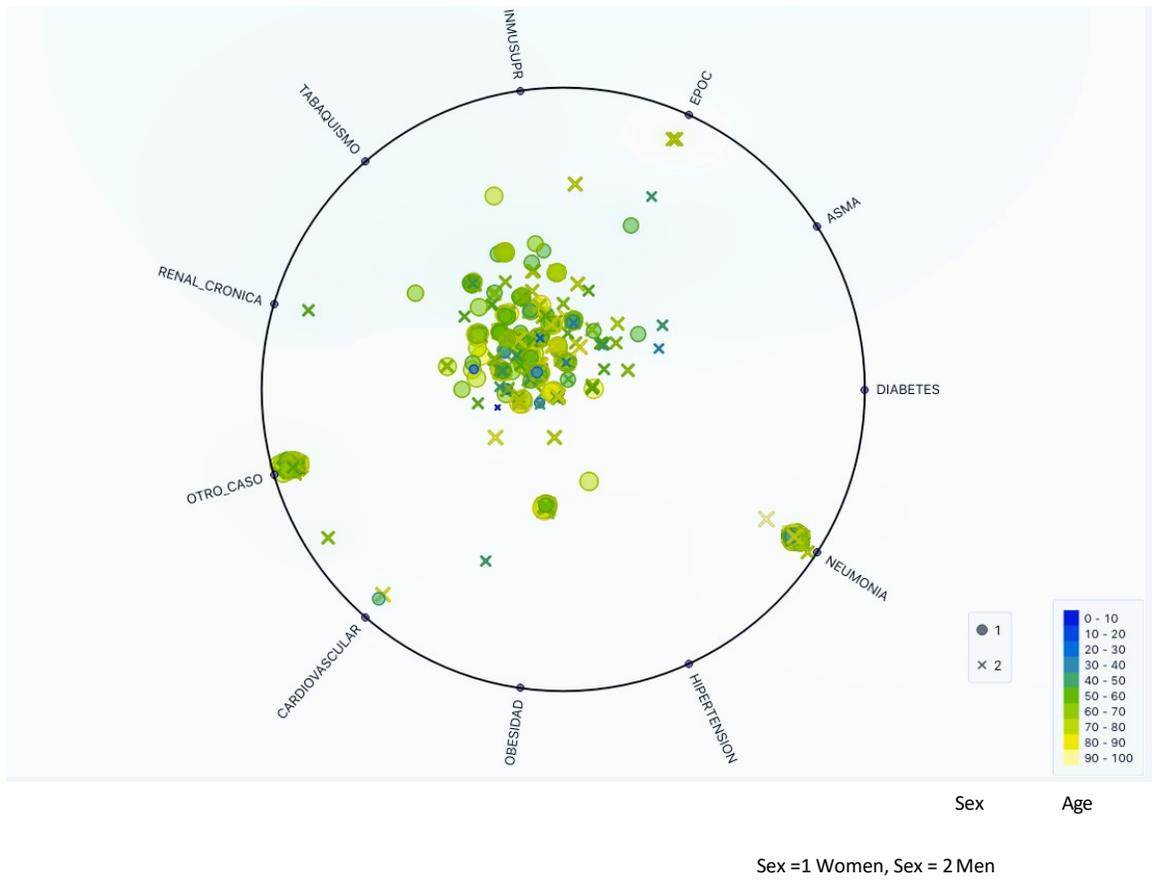

Sex =1 Women, Sex = 2 Men

Source: Own elaboration with Government of Information from the Mexico. Health Secretary. Epidemiological Surveillance System for Viral Respiratory Diseases as of August 1, 2020

**Declaration of Competing Interest**

The authors declare that they have no known competing financial interests or personal relationships which have, or could be perceived to have, influenced the work reported in this article.